\documentclass{epl}

\title{Harper-Hofstadter problem for 2D electron gas
with ${\bf k}$-linear Rashba spin-orbit coupling}
\author{V. Ya. Demikhovskii\inst{1} \and A. A. Perov\inst{1}}

\institute{
  \inst{1} Nizhny Novgorod State University - Gagarin ave., 23, Nizhny
  Novgorod 603950, Russian Federation}

\begin{document}

\maketitle

\begin{abstract}
The Harper-Hofstadter problem for two-dimensional electron gas
with Rashba spin-orbit coupling subject to periodic potential and
perpendicular magnetic field is studied analytically and
numerically. The butterfly-like energy spectrum, spinor wave
functions as well as the spin density and average spin
polarization are calculated for actual parameters of semiconductor
structure. Our calculations show that in two-dimensional electron
gas subject to periodic potential and uniform magnetic field the
effects of energy spectrum splitting caused by large spin-orbit
Rashba coupling can be observed experimentally.
\end{abstract}

\section{Introduction}
The problem of quantum states of an electron subject to both
periodic potential  and homogeneous magnetic field remains to be
actual for several last decades. There are various approaches and
models for periodic potential and different approximations which
are used for investigations of these states (see,
ref.~\cite{b.a}). However, the spin-orbit interaction is usually
excluded from the models as well as the Zeeman term. Such approach
can be justified as long as the amplitude of the periodic
potential is big enough to make the Landau level splitting much
greater than the typical spin-orbit coupling energy and the Zeeman
term. At the same time, under the realistic experimental
conditions with 2D electron gas subject to the potential of
lateral superlattice the potential amplitude $V_0$ can be of the
same order as the spin-orbit (SO) Rashba coupling. For example, in
recent papers~\cite{b.b,b.c} where a pioneering step into
experimental observation of 2D magnetic Bloch states has been
done, the magnitude of $V_0$ was around 1-5 meV. In semiconductors
structures with large SO coupling~\cite{b.5} the typical splitting
energy can be of the same order as $V_0$.

In this paper we study magnetic Bloch states of 2D electrons
subject to both periodic potential of a lateral superlattice and
perpendicular magnetic field under the conditions when the SO
coupling and Zeeman term should be taken into consideration. The
energy band structure is calculated in the magnetic Brillouin zone
(MBZ) and the magnetic Bloch states are constructed. Also, the
spin density distribution in the elementary cell is obtained and
the average spin polarization in a state with given quasimomentum
are calculated. It is shown that the consideration of spin-orbit
coupling is necessary for interpretation of realistic experiments.

\section{Hamiltonian matrix structure, energy band spectrum and wavefunctions}
We consider 2D electron gas with spin-orbit Rashba coupling in a
potential $V(x,y)$ which is periodic in plane with the period $a$,
and in a uniform magnetic field ${\bf H}$ perpendicular to the
plane of the  electrons. The correspondent one-electron
Hamiltonian has the following form:
$$
\hat H=\hat H_0 + V(x,y),\eqno(1)
$$
where $V(x,y)=V(x+na,y+ma)$ is a periodic potential,
$$\hat H_0=({\bf\hat p}-e{\bf A}/c)^2/2m^{\ast}+\frac{\alpha}{\hbar}
\bigg(\hat\sigma_x (\hat p_y-eA_y/c)-\hat\sigma_y\hat p_x\bigg)-
g\mu_B H\hat\sigma_z
$$
is the Rashba Hamiltonian of an electron in uniform magnetic
field~\cite{b.d,b.e}. Here, $\hat p_{x,y}$ are the momentum
operator components, $m^{\ast}$ is the electron effective mass,
${\bf\hat\sigma}$ are the Pauli matrices, $\alpha$ is the
parameter of the SO coupling, $g$ is  the  Zeeman factor, and
$\mu_B$ is the Bohr magneton. We use the Landau gauge in which the
vector potential has the form ${\bf A}=(0,Hx,0)$ and consider the
potential $V(x,y)=V_0(\cos(2\pi x/a)+\cos(2\pi y/a))$. The quantum
states structure of the system under consideration depends
crucially on the parameter $\Phi/\Phi_0=p/q=|e|Ha^2/2\pi\hbar c$
($p$ and $q$ are prime integers) which is the number of flux
quanta per unit cell, and $\Phi_0$ is the flux quanta.

One can express the eigenfunction of Hamiltonian (1) as a set of
Landau wave functions in the presence of the SO
coupling~\cite{b.f}. If we take $p/q$ to be the rational number,
the two-component magnetic Bloch function, analogous to the
one-component magnetic Bloch function of the paper~\cite{b.a}, can
be written in the form
$$
\displaylines{\Psi_{\bf k}(x,y)=\pmatrix{\Psi_{1{\bf k}}(x,y)\cr
\Psi_{2{\bf
k}}(x,y)}=\sum\limits_{S=1}^{\infty}\sum\limits_{n=1}^{p}
\sum\limits_{\ell=-\infty}^{+\infty}{\rm e}^{ik_yy}{\rm
e}^{ik_x(\ell qa+nqa/p)} {\rm e}^{2\pi iy(\ell p+n)/a}\times\cr
\hfill\times\Bigg[A_{0n}({\bf k})\psi_{0n\ell {\bf k}}^{+}(x,y)+
A_{Sn}({\bf k})\psi_{Sn\ell {\bf k}}^{+}(x,y)+B_{S+1,n}({\bf
k})\psi_{S+1,n\ell {\bf k}}^{-}(x,y)\Bigg],\hfill\llap{(2)}\cr}
$$
where  the  spinors $\psi_{0n\ell {\bf
k}}^{+}=\exp(ik_yy)\pmatrix{0\cr \phi_0[\xi_{\ell n}]}$,
$\psi_{Sn\ell {\bf k}}^{+}=\frac{\exp(ik_yy)}
{\sqrt{1+D_S^2}}\pmatrix{D_S\phi_{S-1}[\xi_{\ell n}]\cr
\phi_S[\xi_{\ell n}]}$ and $\psi_{Sn\ell {\bf
k}}^{-}=\frac{\exp(ik_yy)}
{\sqrt{1+D_S^2}}\pmatrix{\phi_{S-1}[\xi_{\ell n}]\cr
-D_S\phi_S[\xi_{\ell n}]}$ correspond to "$+$" and "$-$" branches
of the spectrum of the Hamiltonian $\hat H_0$~\cite{b.f},
$D_S=(\sqrt{2S}\alpha/l_H)/
(E_0^++\sqrt{(E_0^+)^2+2S\alpha^2/l_H^2})$, $\phi_S[\xi]$ is the
simple harmonic oscillator functions, $l_H=c\hbar/|e|H$  is the
magnetic length, $E_0^+=\hbar\omega_c/2+g\mu_BH$ and
$\omega_c=|e|H/m^{\ast}c$ is the cyclotron frequency. Here quantum
numbers $S=1,2,3,\ldots$ characterize the pair of "$+$" and "$-$"
states
$E_S^{\pm}=S\hbar\omega_c\pm\sqrt{(E_0^+)^2+2S\alpha^2/l_H^2}$ of
the unperturbed Hamiltonian $\hat H_0$, $\xi_{\ell n}=(x-x_0-\ell
qa-nqa/p)/l_h$, $x_0=c\hbar k_y/|e|H$.

Note  that the spinor wave function (2) is the eigenfunction of
both the Hamiltonian (1) and operator of magnetic translation and
therefore it have to obey the following Bloch-Peierls conditions
$$
\Psi_{\bf k}(x+qa,y+a)=\Psi_{\bf
k}(x,y)\exp(ik_xqa)\exp(ik_ya)\exp(2\pi iy/a),
$$
where ${\bf k}$ is quasimomentum defined in the MBZ
$$
-\pi/qa\le k_x\le\pi/qa,\quad -\pi/a\le k_y\le\pi/a.
$$
So, the magnetic Brillouin zone is the same as for charged
spinless particle. At the limit of high magnetic fields the
functions $\psi_{Sn\ell {\bf k}}^+$ and $\psi_{Sn\ell {\bf k}}^-$
are proportional to eigenspinors of Pauli operator $\hat\sigma_z$:
$\pmatrix{1\cr 0}$ and $\pmatrix{0\cr 1}$.

Substituting eq.(2) in the Schr\"odinger equation $\hat
H\Psi=E\Psi$ we come to the infinite system of linear equations
for coefficients $A_{Sn}({\bf k})$ and  $B_{Sn}({\bf k})$. In the
case when periodic potential amplitude and spin-orbit coupling
energy have the same order and the inequality $\Delta E_{SO}\simeq
V_0\le \hbar\omega_c$ takes place the system of linear equations
can be reduced to the system of infinite number of uncoupled
groups of $2p$ equations. Each group describes the magnetic Bloch
states formed from the states with energies $E_S^+$ and
$E_{S+1}^-$ of the single Landau level split by SO interaction.

In this approximation the system of $2p$ linear equations attached
to the quantum number $S$ is defined by the following Hamiltonian
matrix
$$
H_{nn\prime}^{SS\prime}=\pmatrix{G_1&M&0&\ldots&0&M^{\ast}&F_1&J&0&
\ldots&0&T^{\ast}\cr
M^{\ast}&G_2&M&0&\ldots&0&T&F_2&J&0&\ldots&0\cr
\ldots&\ldots&\ldots&\ldots&\ldots&\ldots&\ldots&\ldots&\ldots&\ldots&\ldots&\ldots\cr
M^{\ast}&0&\ldots&0&M^{\ast}&G_p&J^{\ast}&0&\ldots&0&T&F_p\cr
F_1^{\ast}&T^{\ast}&0&\ldots&0&J&U_1&N&0&\ldots&0&N^{\ast}\cr
J^{\ast}&F_2^{\ast}&T^{\ast}&0&\ldots&0&N^{\ast}&U_2&N&0&\ldots&0\cr
\ldots&\ldots&\ldots&\ldots&\ldots&\ldots&\ldots&\ldots&\ldots&\ldots&\ldots&\ldots\cr
T&0&\ldots&0&J&F_p^{\ast}&N^{\ast}&0&\ldots&0&N^{\ast}&U_p}\eqno(3)
$$
which  has  the block structure. In eq.(3) the matrix elements are
defined as follows
$$
\displaylines{G_n=E_S^++V_0{\rm e}^{-\pi q/2p}\cos(2\pi x_0/a+2\pi
nq/p) [D_S^2L_{S-1}^0(\pi q/p)+L_S^0(\pi
q/p)]\frac{1}{1+D_S^2},\cr M=\frac{V_0}{2}\exp(ik_xqa/p){\rm
e}^{-\pi q/2p}[D_S^2L_{S-1}^0(\pi q/p)+ L_S^0(\pi
q/p)]\frac{1}{1+D_S^2},\hfill\cr U_n=E_{S+1}^-+V_0{\rm e}^{-\pi
q/2p}\cos(2\pi x_0/a+2\pi nq/p) [D_{S+1}^2L_{S+1}^0(\pi
q/p)+L_S^0(\pi q/p)]\frac{1}{1+D_{S+1}^2},\hfill\cr
N=\frac{V_0}{2}{\rm e}^{ik_xqa/p}{\rm e}^{-\pi
q/2p}[D_{S+1}^2L_{S+1}^0(\pi q/p)+ L_S^0(\pi
q/p)]\frac{1}{1+D_{S+1}^2},\hfill\llap{(4)}\cr F_n=V_0{\rm
e}^{-\pi \frac{q}{2p}}\sin(2\pi\frac{x_0}{a}+2\pi\frac{nq}{p})
\Bigg[\frac{D_{S+1}}{\sqrt{S+1}}L_S^1(\pi
q/p)-\frac{D_S}{\sqrt{S}}L_{S-1}^1(\pi q/p)\Bigg]
\frac{1}{(1+D_S^2)(1+D_{S+1}^2)},\hfill\cr
J=\frac{V_0}{\sqrt{2}}{\rm e}^{i\frac{k_xqa}{p}}{\rm e}^{-\pi
\frac{q}{2p}}\sqrt{\frac{\pi q}{2p}}
\Bigg[\frac{D_{S+1}}{\sqrt{S+1}}L_S^1(\pi
q/p)-\frac{D_S}{\sqrt{S}}L_{S-1}^1(\pi q/p)\Bigg]
\frac{1}{(1+D_S^2)(1+D_{S+1}^2)},\hfill\cr
T=-\frac{V_0}{\sqrt{2}}{\rm e}^{-i\frac{k_xqa}{p}}{\rm e}^{-\pi
\frac{q}{2p}}\sqrt{\frac{\pi q}{2p}}
\Bigg[\frac{D_{S+1}}{\sqrt{S+1}}L_S^1(\pi
q/p)-\frac{D_S}{\sqrt{S}}L_{S-1}^1(\pi q/p)\Bigg]
\frac{1}{(1+D_S^2)(1+D_{S+1}^2)},\hfill\cr}
$$
Here $L_S^K(z)$ is the Laguerre polynomial.

This system has a form of generalized Harper's equation. Matrix
elements $G_n$ and $M$ in the left upper block of eq.(3)
characterize the interaction between different states of the "$+$"
branch and elements $U_n$ and $N$ in right low block are due to
the interaction between states of the "$-$" branch. The right
upper and left low blocks with elements $F_n,\, J$ and $T$
describe the transition between states of the neighbor "$+$" and
"$-$" branches with index $S$. Matrix elements in these four
blocks are periodic in $n$ with period $p$. So, each block has the
size $p\times p$.

Numerical calculations of energy spectrum and wave functions were
carried out for the actual parameters which approximately
correspond to the structure used in the experiment. In our
calculations we used the following parameters of 2D electron gas:
$V_0=1\, meV$, effective mass $m^{\ast}=0.05\, m_0$,
$g=g_{InAs}=-2.0$, $\alpha=5\cdot 10^{-11}\, eV\cdot m$. The
lattice period was chosen as $a=60\, nm$ and $V_0=1\, meV$. Under
these parameters the corresponding spectrum structure in magnetic
field becomes well resolved and SO splitting is larger than Zeeman
splitting at $p/q\le 5$. The chosen parameters approximately
correspond to the InAs quantum well where the Rashba SO constant
can reach the maximal value $5\cdot 10^{-11}\, eV\cdot m$ due to
electron wave function penetration into barrier layer~\cite{b.5}.

In fig.1 we show the electron energy levels for the state ${\bf
k}=0$ versus the magnetic flux number. The big dots define the
position of Landau levels $E_S^{\pm}$ in the absence of periodic
potential. The insert demonstrates the magnetic subbands at $p/q$
equal to $3/1$, $4/1$ and $5/1$ for ${\bf k}$ located in the
magnetic Brillouin zone. The upper arrow marks the lowest magnetic
subband for which the dispersion law $E_1({\bf k})$ is plotted in
fig.2. Note that the spectrum of magnetic subbands has the
symmetry of the $C_{4V}$ group, as it should be. As one can see
from fig.1 the periodic potential forms the spectra which resemble
the Hofstadter butterflies at the region $1\le p/q \le 2$. For
$p/q\ge 3$ the magnetic subbands split by SO coupling and periodic
potential are not overlapped. So, at magnetic fields corresponding
to $p/q\ge 3$ the unperturbed levels (fat dots in fig.1) are
grouped into pairs and, therefore, our two-branch approximation is
appropriate.

\begin{figure}
\onefigure[scale=0.7]{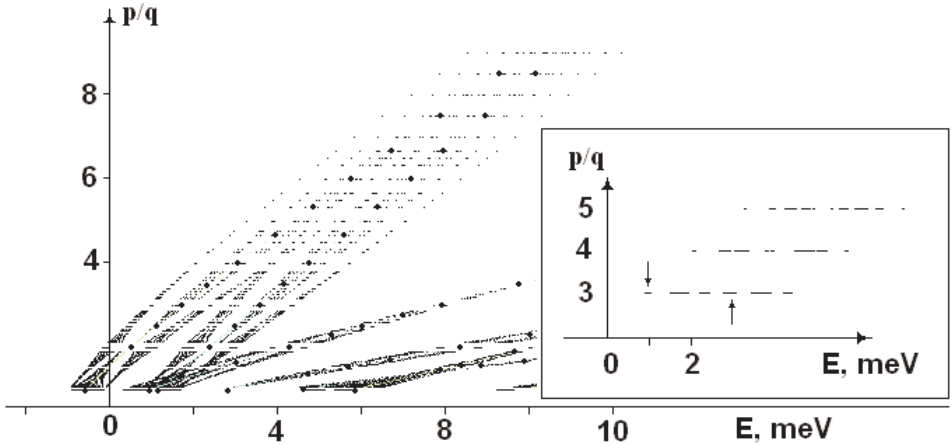} \caption{Position of energy
subbands versus magnetic flux numbers at the parameters indicated
in the text. The big dots show "$+$" and "$-$" Landau level
positions in the absence of periodic potential. The insert shows
$2p$ energy subbands for $p/q=3/1,\, 4/1,\, 5/1$.} \label{f.1}
\end{figure}

\begin{figure}
\onefigure[scale=0.5]{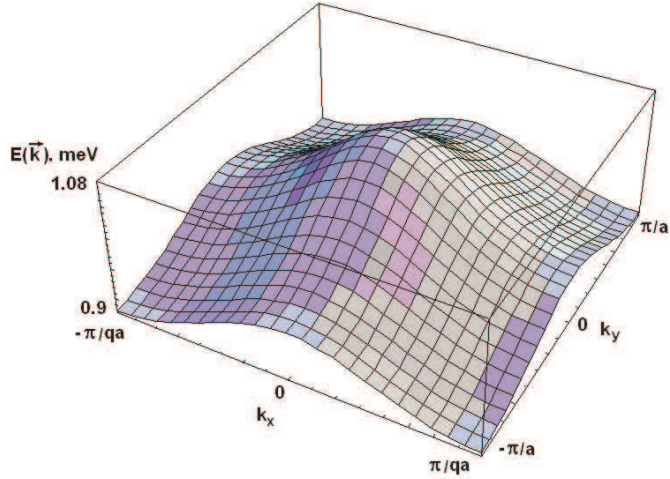} \caption{Electron energy versus
quasimomentum in magnetic Brillouin zone.} \label{f.2}
\end{figure}

To compare the effects of level splitting due to the periodic
potential as well as Zeeman effect and SO coupling we calculated
the energy spectra formed from the lowest pair of Landau levels
(see fig.3). In fig.3a we show the Hofstadter-like spectrum in the
absence of Zeeman and SO interactions. Here, all levels are
two-fold degenerate. At the presence of Zeeman interaction
(fig.3b) the degeneracy is lifted and the correspondent spectrum
is the superposition of two Hofstadter-like butterflies. Fig.3c
demonstrates the splitting due to the SO Rashba coupling. So,
under the parameters and magnetic fields indicated above the SO
interaction is large enough to form two non-overlapped and
non-degenerate groups of magnetic subbands.

\begin{figure}
\onefigure[scale=0.7]{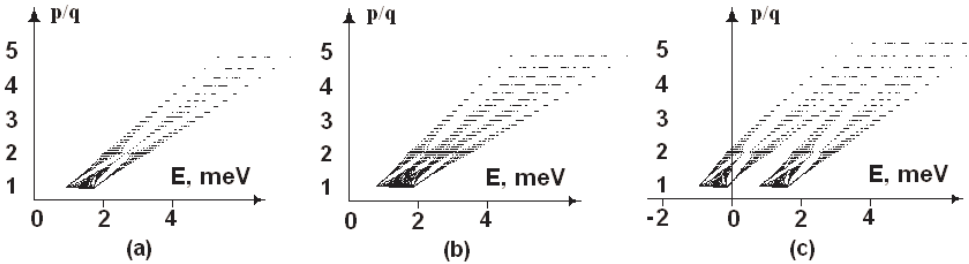} \caption{Energy spectra in the
absence of SO Rashba coupling and Zeeman effect (a), in the
presence of Zeeman splitting only (b), and when all of the
contributions to level splitting are present (c). The parameters
are the same as in fig.1.} \label{f.3}
\end{figure}

We have calculated also the electron density $|\Psi_{\bf
k}(x,y)|^2=|\Psi_{1{\bf k}}(x,y)|^2+|\Psi_{2{\bf k}}(x,y)|^2$ at
${\bf k}=0$ for the fourth magnetic subband labelled by low arrow
at the insert in fig.1. At the center of MBZ it has the symmetry
of the $C_{4V}$ group. Note, that for the state ${\bf k}\ne 0$ the
electron density does not possess the symmetry of the lattice.

\begin{figure}
\onefigure[scale=0.5]{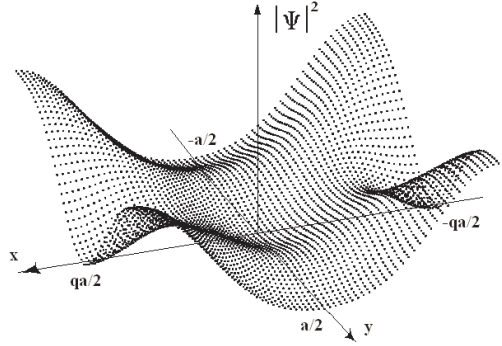} \caption{The electron density
$|\Psi_{\bf k}(x,y)|^2= |\Psi_{1{\bf k}}(x,y)|^2+|\Psi_{2{\bf
k}}(x,y)|^2$ at ${\bf k}=0$ for the fourth magnetic subband.}
\label{f.4}
\end{figure}

\section{Spin density and average spin polarization}
In the 2D electron gas with spin-orbit coupling at the presence of
the periodic potential and perpendicular magnetic field the
non-trivial spin structure appears in each of magnetic energy
bands. Such a structure at a certain ${\bf k}$ can be
characterized by the vector field of the spin density in
coordinate space
$$
S_{i{\bf k}}(x,y)=\Psi_{\bf k}^+({\bf r})\hat\sigma_i\Psi_{\bf
k}({\bf r}), \,\,\,(i=x,y).
$$
In addition, spin states can be characterized by the average spin
distribution in ${\bf k}$ space
$$
S_i({\bf k})=<S_{i{\bf k}}(x,y)>=<\Psi_{\bf k}^+({\bf
r})\hat\sigma_i\Psi_{\bf k} ({\bf r})>,
$$
where the brackets mean the integration on $x$ and $y$ over the
magnetic unit cell.

\begin{figure}
\onefigure[scale=0.6]{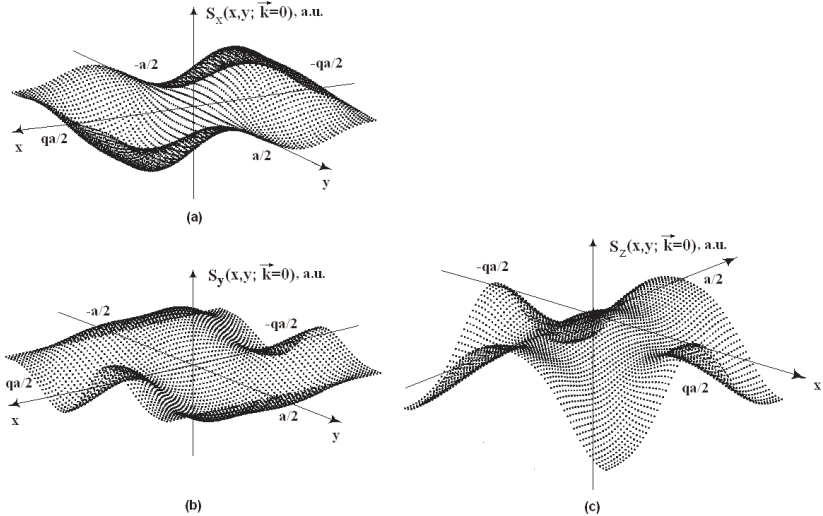} \caption{Electron spin densities
in a unit cell of the lateral superlattice.} \label{f.5}
\end{figure}

In fig.5 we show the spin density distribution calculated in the
${\bf r}$-space for the square superlattice with parameters
indicated above. All results correspond to the fourth magnetic
energy band shown at the insert in fig.1 for the state ${\bf
k}=0$. The $S_{x{\bf k}=0}(x,y)$ and $S_{y{\bf k}=0}(x,y)$
densities have the $C_s$ symmetry, but the $S_{z{\bf k}=0}(x,y)$
component has the $C_{4V}$ symmetry. The integration of the spin
densities over the magnetic unit cell leads to the zero values of
$S_x({\bf k})$ and $S_y({\bf k})$ at ${\bf k}=0$ (see fig.6a). The
integration of $S_{z{\bf k}=0}(x,y)$ over the magnetic unit cell
results to the positive value of the average spin projection along
magnetic field.

\begin{figure}
\onefigure[scale=0.7]{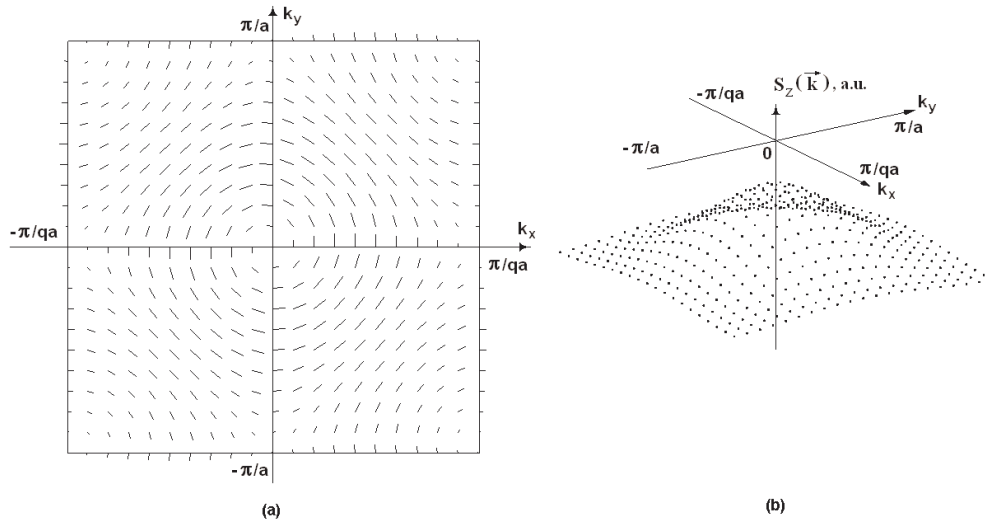} \caption{(a) Distribution of
average spin polarization $S_x({\bf k})$ and $S_y({\bf k})$ in the
fourth magnetic subband attached to the two lowest neighbor
branches. (b) $S_z({\bf k})$ in the fourth magnetic energy
subband.} \label{f.6}
\end{figure}

In fig.6 we show the average spin polarization in the MBZ for the
fourth band at the magnetic flux $p/q=3/1$ (see fig.1). Fig.6a
demonstrates the distribution of the components $S_x({\bf k})$ and
$S_y({\bf k})$ and fig.6b shows the $S_z({\bf k})$ component. The
distribution in fig.6a has the vortex structure with the $C_2$
symmetry. The vortex centers are located at the center and at the
corners of MBZ. These two vortices have opposite directions of
rotation. In these points the average lateral spin components are
smaller than perpendicular component $S_z({\bf k})$. For the state
${\bf k}=0$ such a behavior is due to the small value of SO
coupling in comparison with Zeeman effect. At the corners of MBZ
the SO coupling is effectively smaller than the spin splitting in
a magnetic field as a result of state mixing at the magnetic
Brillouin zone edges. Here, in accordance with normalizing
condition the $|S_z({\bf k})|$ component has the local maxima. We
would like to note here that these spin vortices define the Berry
phases and, respectively, the spin Hall conductance. Due to the
same fact of state mixing by periodic potential the SO coupling
effectively vanishes at the points where the edges of MBZ cross
the $k_x=0$ and $k_y=0$ lines. Finally, note that the total
magnetic moment of 2D electron gas which completely occupies this
zone is positive.

\section{Conclusion}
We have considered the effects of spin-orbit coupling on the
quantum states of two-dimensional electron gas at the presence of
periodic potential and uniform magnetic field. The Hofstadter
butterfly-like energy band spectrum and electron density in the
unit cell was analyzed. We demonstrate that the Zeeman interaction
and the spin-orbit coupling form a complex and intricate spin
periodic structure in the field of the lateral superlattice. The
electron spin densities $S_{x{\bf k}}(x,y)$, $S_{y{\bf k}}(x,y)$
and $S_{z{\bf k}}(x,y)$ as well as average spin polarizations
$S_x({\bf k})$, $S_y({\bf k})$ and $S_z({\bf k})$ for the states
with different quasimomentum ${\bf k}$ defined in magnetic
Brillouin zone were calculated and the role of three contribution
to the spin orientation was investigated. We think that our
results will be useful for analyzing the data of real experiments
on the hunting of Hofstadter butterfly in two-dimensional electron
gas.

\acknowledgments This work was supported by the program of Russian
Ministry of Education and Science:  Development of scientific
potential of High School (project 2.1.1.2363), grant of Russian
Foundation of Basic Research (no. 06-02-17189) and grant of the
President of Russian Federation (MK-5165.2006.2).

\end{document}